\documentclass[12pt]{article}
\usepackage{graphicx}
\usepackage{epstopdf}
\usepackage{cite}

\makeatletter

\makeatletter

\newcommand{\be}{\begin{equation}}
\newcommand{\ee}{\end{equation}}
\newcommand{\beq}{\begin{eqnarray}}
\newcommand{\eeq}{\end{eqnarray}}

\def\rf#1{(\ref{#1})}

\def\[{\left [}
\def\]{\right ]}
\def\({\left (}
\def\){\right )}

\def\b{\beta}

\def\S{{\bf S}}
\def\A5S5{{\rm AdS}_5 \times \S^5}

\def\nc{non-commutative}
\def\lnc{lightlike \nc}

\title{{\bf Causally pathological spacetimes are physically relevant}}

\author{Veronika E. Hubeny$^{a,b,}$, Mukund Rangamani$^{a,b}$ and 
Simon F. Ross$^a$ 
\footnote{veronika.hubeny, mukund.rangamani, s.f.ross@durham.ac.uk}\\ \\
\small \sl $^a$Centre for Particle Theory \& Department of
Mathematical Sciences, 
\\[-1.5mm]
\small \sl Science Laboratories, South Road, Durham DH1 3LE, United Kingdom\\ 
\small\sl $^b$Department of Physics \& Theoretical Physics Group,
LBNL,\\[-1.5mm] 
\small\sl University of California, Berkeley, CA
94720, USA\\ }

\begin{document}
\setlength{\baselineskip}{16pt}
\begin{titlepage}
\maketitle
\begin{picture}(0,0)(0,0)
\put(335,340){gr-qc/0504013}
\put(335,325){DCPT-05/16}
\put(335,310){UCB-PTH-05/08}
\put(335,295){LBNL-57391}
\end{picture}
\vspace{-36pt}
\begin{abstract}
We argue that in the context of string theory, the usual restriction
to globally hyperbolic spacetimes should be considerably relaxed. We
exhibit an example of a spacetime which only satisfies the causal
condition, and so is arbitrarily close to admitting closed causal
curves, but which has a well-behaved dual description, free of paradoxes.
\end{abstract}
\thispagestyle{empty}
\setcounter{page}{0}
\end{titlepage}


A striking feature of general relativity is that it has solutions with
nontrivial, and sometimes even pathological, causal properties.
Although the local light cone is that of special relativity, the global
causal properties can be completely different; spacetimes can have
closed timelike curves. Observers in such a spacetime can return to
the same point in spacetime, leading to the famous ``grandfather
paradox''.

Time machines have continued to fascinate the science-fiction writer
and the student of general relativity alike. Although many solutions
with closed timelike curves have been found, it is much more difficult
to construct examples where a time machine develops from causally
regular initial conditions. In the Kerr solution, for example, there
is a region of closed timelike curves inside the inner horizon, but
this horizon is believed to be unstable due to an infinite
blueshift. This has led to the celebrated chronology protection
conjecture~\cite{Hawking:1991nk}, which asserts the impossibility of
forming a time machine. This has remained an open question, and a full
resolution may require a deeper understanding of quantum gravity. In
practical investigations, however (for example, in numerical
relativity, and canonical approaches to quantum gravity), attention is
commonly restricted to globally hyperbolic spacetimes, which admit a
global Cauchy surface. 

String theory is a candidate theory of quantum gravity, and it has now
been developed to the point where we have a well-studied proposal for a
nonperturbative and background independent description of a certain
class of spacetimes, via the AdS/CFT
dualities~\cite{Aharony:1999ti}. It is interesting to ask what this
proposal has to say about the question of causality violations. The
first thing to note is that the usual restriction to globally
hyperbolic spacetimes must be relaxed, as Anti-de Sitter space, the
paradigmatic example of the duality, is not globally hyperbolic.

How far do we need to relax? A rich hierarchy of causality properties,
weaker than global hyperbolicity, has been defined in the past.  In
order of increasing strength, the most common conditions are: causal
$<$ weakly distinguishing $<$ (future and past) distinguishing $<$
strongly causal $<$ stably causal $<$ globally hyperbolic.  (For a
review, see~\cite{Hawking:1973ab}.) The most natural criterion would
be to consider stably causal spacetimes. A spacetime is {\it
stably causal} if it remains causal under any sufficiently
small deformation of the metric. In a theory of quantum gravity, where
the metric is subject to quantum fluctuations, one would therefore expect
that stable causality should constitute the minimal requirement needed to
avoid possible pathologies. Also, spacetimes which are causal but not
stably causal must, by definition, form a subset of measure zero in
the space of continuous metrics, and until recently, few physically
interesting examples of such spacetimes were known. In standard texts
such as~\cite{Hawking:1973ab}, examples of such spacetimes are
constructed by the rather unphysical method of excising points from
the spacetime manifold.

Indeed, Anti-de Sitter space is stably causal. The failure of global
hyperbolicity is a result of the asymptotic behaviour: it has a
timelike conformal boundary. This is important, as it implies that
once we specify appropriate boundary conditions on this conformal
boundary, the spacetime will have a well-defined initial value
problem, defining `Cauchy' evolution from an initial slice. There is a
Cauchy horizon associated with each Cauchy surface in this spacetime,
but unlike in the example of the Kerr spacetime, it is not a surface
of infinite blueshift. Hence there is no natural mechanism to render
the physical part of the spacetime globally hyperbolic. The failure of
global hyperbolicity is thereby closely tied to the dual CFT, whose
definition is linked to the asymptotic boundary conditions in
spacetime.

Another interesting class of backgrounds which are stably causal but
not globally hyperbolic are plane waves. The story is similar
for these solutions. It was observed by Penrose~\cite{Penrose:1965rx}
that they are not globally hyperbolic spacetimes, and more recently,
they were shown to be stably causal~\cite{Hubeny:2003sj}. The failure
of global hyperbolicity is again associated with the asymptotic
behaviour of the spacetime. A family of plane wave spacetimes has been
shown to have a one-dimensional null boundary~\cite{Berenstein:2002sa,
Marolf:2002ye}, which is both to the past and future of points in the
interior of the spacetime.  Given boundary conditions on this
conformal boundary, a well-defined initial data problem should
exist. Certain plane waves have also been related to dual field theory
descriptions through the BMN correspondence~\cite{Berenstein:2002jq},
which is obtained by considering the plane wave that arises in the
Penrose limit of an asymptotically AdS spacetime.

The previous two examples suggest a picture where we relax the
condition of global hyperbolicity only slightly, considering stably
causal spacetimes which, once we specify appropriate asymptotic
boundary conditions, still have a well-defined initial value
problem. 

However, we now wish to argue that we actually need to relax it much 
more radically, and extend our criterion to include causal spacetimes 
which are not even distinguishing. (A spacetime is {\it causal} if it does not contain
any closed causal curves, and it is {\it distinguishing} if distinct
points have distinct causal past and future sets.)

The first evidence that we should allow ourselves to consider
spacetimes with such weak causal properties comes from the recent
construction of simple uncontrived examples. It was pointed out
in~\cite{Flores:2004dr, Hubeny:2003sj} that for 
pp-waves\footnote{pp-waves are generalizations of
plane waves, which still have a covariantly constant
null Killing field, but are less symmetric.}, whose metric can be written as
$$ds^2 = -2 \, du \, dv - F(u, x^i) \, du^2 + dx^i \, dx^i, $$ 
if the function $F(u, x^i)$ grows faster than quadratically in $x^i$
for some $x^i$, the spacetime is actually {\it not} stably causal; in
fact, it is not even distinguishing!  Indeed, such pp-waves are rather
spectacularly non-distinguishing, as {\it all} spacetime points with
the same $u$ coordinate (say $u = u_0$) have the same past and future,
consisting of all points with $u<u_0$ and $u>u_0$,
respectively. In~\cite{Maldacena:2002fy} it was shown that some
examples, where the function $F(u, x^i)$ grew exponentially at large
$x^i$ for some $x^i$, could be described in string theory, via a
world-sheet conformal field theory with no obvious pathologies.

\begin{figure}[htbp]
\begin{center}
\includegraphics[width=5.5in]{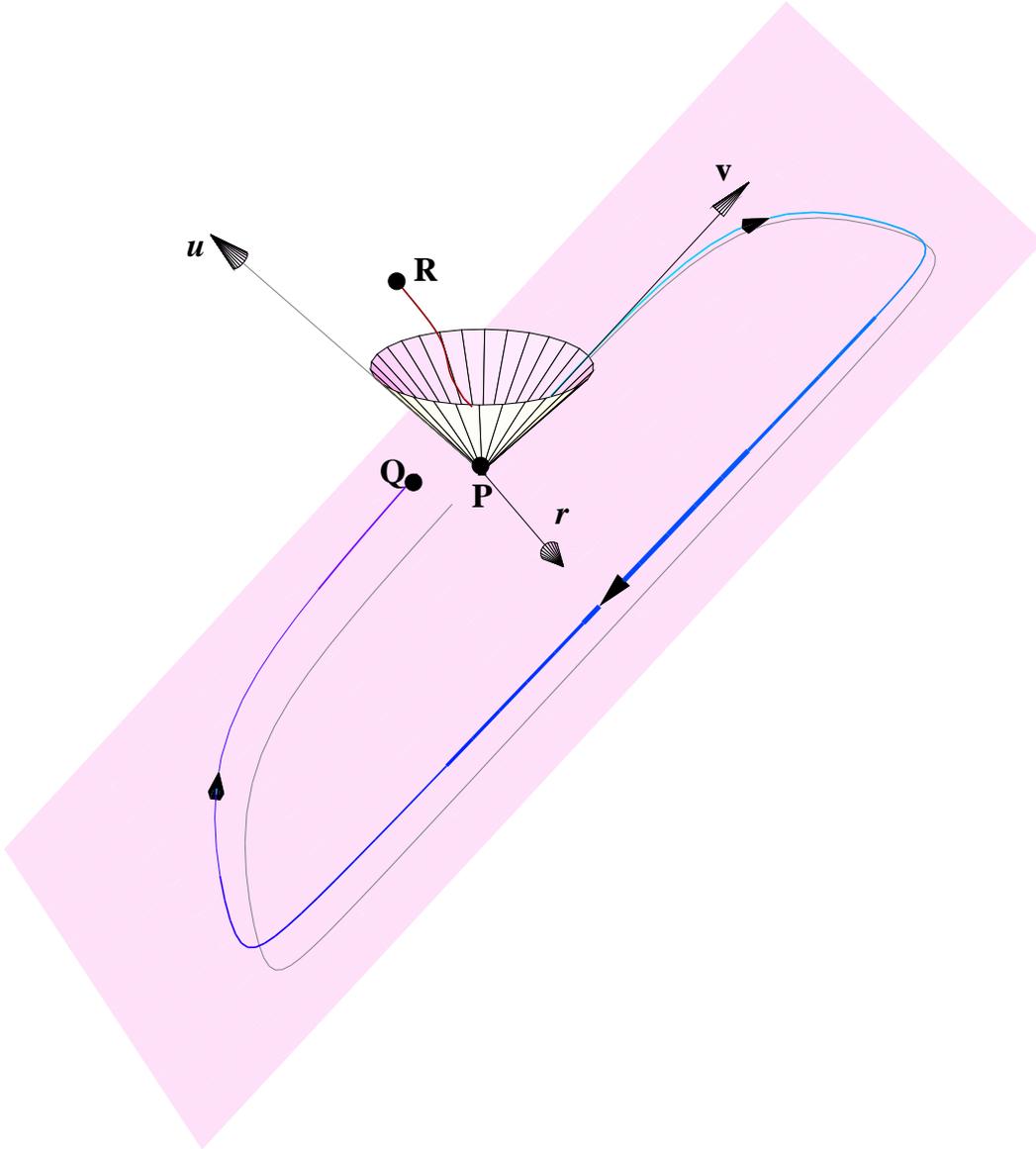}
\caption{{\small Light cone structure in the non-distinguishing spacetime 
\rf{ndgeo}. From a point ${\bf P}$ in the spacetime, it is possible to 
reach point ${\bf R}$ by a usual causal curve. To reach point ${\bf Q}$
(which is ``close'' to ${\bf  P}$) we need to go out towards the asymptotic 
region and then loop back. Such a curve is drawn above along with its 
projection onto a constant $u= u_{\bf P}$ plane. By going out further in $r$ 
we can engineer curves to return arbitrarily close to our starting point. 
Existence of these causal curves 
implies that the causal future of ${\bf P}$ is the part of the spacetime above 
the plane of constant $u$ drawn in the figure. Contrast this with the 
flat space light cone of the point.   }}
\label{nondistlc}
\end{center}
\end{figure}

However, lacking a full non-perturbative description, one might still
question whether such spacetimes can be physically relevant, as they
are arbitrarily close to developing closed timelike curves. A new
example, which is related to a field theory description via an AdS/CFT
type duality, provides strong evidence that in fact they are
physically relevant. The metric may be written as
\begin{eqnarray}
 ds^2 &=&  {r^2 } \, \( -2 \, du \, dv  - \b^2 \, r^4  \, du^2+ 
 dx^2 + dy^2 +  {dr^2 \over r^4} + {1\over r^2} \, d\Omega_5^2 \)  
 \nonumber \\
&=&  {1\over z^2} \, \(-2 \, du \, dv  - {\b^2 \over z^4}  \, du^2
+ dx^2 + dy^2 +  dz^2 +  z^2 \, d\Omega_5^2 \) \ .
\label{ndgeo}
\end{eqnarray}
The second line, included to explicitly demonstrate the spacetime being conformal 
to a pp-wave, is obtained by $ z = 1/r$.
The  asymptotic behaviour of the metric as $r \to \infty$ is responsible 
for its non-distinguishingness. Note that for $\b =0$ the metric is that 
of $\A5S5$ in Poincar\'e coordinates. In Fig.~\ref{nondistlc} we illustrate
some causal properties of the spacetime \rf{ndgeo}.

This geometry can be obtained by a solution generating transformation
from the Anti-de Sitter spacetime involved in the AdS/CFT
correspondence (for details, see~\cite{us}). One can consider the
action of this transformation in the field theory variables, to show
that this geometry is also related to a field theory ``living on the
boundary" --- in the present case the dual is a \lnc\ field
theory. This is a non-local theory, but it nevertheless obeys all the
usual properties of a quantum field theory and hence serves as a
non-perturbative description of quantum gravity in the background
\rf{ndgeo}. We take the view that any spacetime which is dual to
a well-defined quantum field theory must itself be sensible (physically
relevant). The field theory description guarantees that the causal
pathologies cannot lead to paradoxes. One can in fact show that the
non-distinguishing nature of the spacetime is a necessary consequence
of the non-local interactions of the \lnc\ field theory; the
micro-causality conditions in the field theory force this unusual
causal behaviour on the dual spacetime!

Note that in \rf{ndgeo}, the violation of stronger causality
conditions comes again from the asymptotic region of spacetime, at $r \to
\infty$. Since the relation to the field theory fixes the asymptotic
boundary conditions, the metric should not be thought of as freely
fluctuating in this $r \to \infty$ region. Hence, it will not suffer from
any problems associated with closed timelike curves, even though the
field theory really corresponds to treating the metric quantum
mechanically. (As a result, we do not regard this example as evidence that
we should consider spacetimes with closed timelike curves as physical.)

Thus, in the AdS/CFT context, we are driven to relax the causality
requirement as far as possible: not only is global hyperbolicity too
stringent a requirement, but even the weak distinguishing property
appears to be too constraining!  There exist spacetimes which are
arbitrarily close to admitting closed causal curves but which are still
dual to  sensible quantum field theories, free of paradoxes. We
therefore conclude that spacetimes wherein the asymptotic
behaviour leads to violations of the stronger causality conditions 
may play an important role. 

\section*{Acknowledgements}

VH and MR are supported by the  funds from the Berkeley Center for
Theoretical Physics, DOE grant DE-AC03-76SF00098 and
the NSF grant PHY-0098840. SFR is supported by the EPSRC Advanced 
Fellowship.

\vskip1cm


\providecommand{\href}[2]{#2}\begingroup\raggedright\endgroup

\end{document}